\documentclass[twocolumn,tightenlines,pra,showpacs]{revtex4}

\usepackage{amssymb}
\usepackage{amsmath}
\usepackage{colordvi}
\usepackage{amsthm}
\usepackage{epsfig}
\usepackage{graphicx}
\usepackage{subfigure}
\usepackage{color}
\usepackage{hyperref}
\usepackage{url}

\begin{document}

\title{Two-component Bose-Hubbard model in an array of cavity polaritons}
\author{Yong-Chang Zhang$^{1,2}$}
\author{Xiang-Fa Zhou$^{1,2}$}
\author{Xingxiang Zhou$^{1,2}$}
\author{Guang-Can Guo$^{1,2}$}

\author{Han Pu$^{3,4}$}
\email{hpu@rice.edu}

\author{Zheng-Wei Zhou$^{1,2}$}
\email{zwzhou@ustc.edu.cn}
\affiliation{$^1${Key Laboratory of Quantum Information,
University of Science and Technology of China, CAS, Hefei, Anhui
230026, People's Republic of China}\\
$^2${Synergetic Innovation Center of Quantum Information and Quantum Physics, University of Science and Technology of China, Hefei, Anhui 230026, China}\\
$^3${Department of Physics and Astronomy, Rice University, Houston, TX 77005, USA}\\
$^4${Center for Cold Atom Physics, Chinese Academy of Sciences, Wuhan 430071, China}}

\begin{abstract}
We propose a scheme which can realize an extended two-component Bose-Hubbard model using polaritons confined in an array of optical cavities. In addition to the density-dependent interactions, this model also contains nonlinear coupling terms between the two components of the polariton. Using a mean-field calculation, we obtain the phase diagram which shows how these terms affect the transition between the Mott insulator and the superfluid phase. In addition, we employ both a perturbation approach and an exact diagonalization method to gain more insights into the phase diagram.
\end{abstract}

\maketitle
\section{Introduction}

Over the past decade, with the development of the technique of the optical lattices, one could manipulate cold atoms in the periodic potentials, which serve as an ideal platform to explore various many-body models in condensed matter physics \cite{tcbhm1,olbhm1,olbhm2}. As one of the important fundamental models in many-body physics, the Bose-Hubbard model(BHM) has attracted lots of attention since it was first introduced in 1963 \cite{bhm}. In the single component BHM, there exists a quantum phase transition between the Mott insulator(MI) phase and the superfluid (SF) phase \cite{misf1,misf2}. This transition has been explored in recent cold atom experiments \cite{olbhm2}. When this is extended to two components of bosonic modes, the phase diagram becomes more complicated correspondingly \cite{tcbhm2,tcbhm3,tcbhm4}.

The standard single-component BHM only includes the on-site repulsive interaction $U$ and the tunnelling $t$ between nearest neighbours. The interaction term favors the MI phase in which each site hosts a definite number of particles; whereas the tunnelling term favors the superfluid phase where the particles are delocalized. The competition between these two terms drives a quantum phase transition between these two phases. For the two-component BHM, besides the intra-species interaction $U_i$ and the tunnelling $t_i$ for each component (where $i=1,2$ labels the components), there is the additional inter-species interaction between the two components \cite{tcbhm1}. In recent years, a lot of attention has been paid to the extended single (two or more) component BHM, and one kind of the extended BHM has additional tunnelling between the next-nearest neighbours or long range interactions \cite{ebhm1,ebhm2,ebhm9,ebhm10,ebhm11}, while another kind has additional nonlinear coupling between different components \cite{ebhm4,ebhm5,ebhm6,ebhm7,TRS,ebhm8}. Expectedly, these additional terms in the extended BHM give rise to a more complicated phase diagram \cite{ebhm2,ebhm6,ebhm7,TRS,ebhm8}.

In a previous work \cite{etcbhm1}, we investigated a system of two-component polariton confined in a single optical microcavity. Polariton is a kind of bosonic quasiparticle in the light-matter system, and Bose-Einstein condensation (BEC) of microcavity polaritons have been realized in experiment recently \cite{pbec1,pbec2}. In this model, in addition to the density-density interaction (Kerr nonlinearity), there also exist two types of nonlinear coupling between the polariton components. In the current work, we extend this model by consider a one dimensional array of such a cavity polariton system, which realizes a two-component BHM. We will calculate the phase diagram of the system and focus on how the inter-species interactions and the nonlinear coupling terms will affect the transition between the MI phase and the SF phase.

This paper is organized as follows. In Sec.~II we present the model Hamiltonian that describes an extended two-component Bose-Hubbard model. In Sec.~III, we provide the ground state phase diagram obtained from the mean-field decoupling approach. To gain futher insights and examine the effects of different terms in the Hamiltonian, we present a perturbative analysis in Sec.~IV. In Sec.~V, we give a beyond-mean-field exact diagonalization calculation. In particular, we calculate the number fluctuation per cavity, and establish connections between this result with the mean-field result presented earlier. Finally, a summary is presented in Sec.~VI. 


\begin{figure}[h]
   \includegraphics[width=0.9\columnwidth]{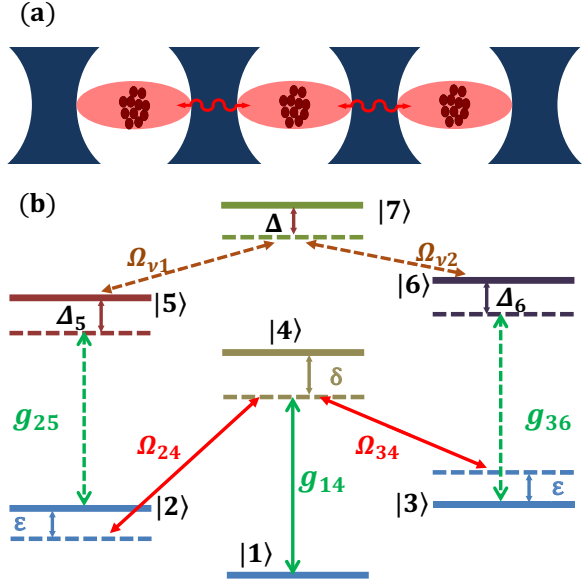}
\caption{(Color online) (a) An illustration of a one-dimension array of optical microcavities. Each cavity host a system of two-component polaritons. (b) The atomic level structure and coupling laser fields that give rise to the two-component polariton system in each cavity. Here $g_{ij}$ is the coupling between the dipole transition ($\vert i \rangle \leftrightarrow \vert j \rangle$) induced by the cavity mode, $\Omega_{ij}$ and $\Omega_{\nu k}$ are the Rabi frequencies of the driving laser fields and driving microwave fields respectively. For details, please see Ref.~\cite{etcbhm1}.}
\label{figmodel}
\end{figure}

\section{Model}
Our underlying system is schematically shown in Fig.~\ref{figmodel}(a). We consider a one-dimensional array of connected optical cavities, with the connection provided by the photon tunneling between neighboring cavities. Insider each cavity, we have an ensemble of bosonic atoms whose level structure is sketched in Fig.~\ref{figmodel}(b). We label the atomic hyperfine states as $|i \rangle$ with $i=1,2,...,7$. Three of these states (states $|1 \rangle$, $|2 \rangle$, and $|3 \rangle$) belong to the electronic ground  manifold, and the other four (states $|4 \rangle$, $|5 \rangle$, $|6 \rangle$, and $|7 \rangle$) belong to the electronic excited manifold. The ground states  $|1 \rangle$, $|2 \rangle$, and $|3 \rangle$ are dipole coupled to the excited states $|4 \rangle$, $|5 \rangle$, and $|6 \rangle$, respectively, by the cavity field, with corresponding coupling strengths $g_{14}$, $g_{25}$ and $g_{36}$. Additionally, states $|2 \rangle$ and $|3 \rangle$ are coupled to $|4 \rangle$ by external {laser fields} with coupling strengths $\Omega_{24}$ and $\Omega_{34}$, respectively. Finally, within the excited manifold, states $|5 \rangle$ and $|6 \rangle$ are coupled to $|7 \rangle$ by microwave fields with corresponding coupling strengths $\Omega_{\nu 1}$ and $\Omega_{\nu 2}$. $\epsilon$, $\delta$, $\Delta$, $\Delta_5$ and $\Delta_6$ are various detunings between the driving field and the corresponding atomic transitions, as labelled in Fig.~\ref{figmodel}(b).

In the limit of weak excitation where the atomic population in the excited levels is negligible, we can construct two polariton modes for each cavity. The corresponding annihilaton operator for the two polariton modes in the $i$th cavity is given by 
\begin{eqnarray*} 
P_{1i}
&=&  \frac{1}{2} 
\left(\frac{g_i}{\omega_i}+ 1\right) S_{2i}+\frac{1}{2} 
\left(\frac{g_i}{\omega_i} - 1 \right) S_{3i}-\frac{\Omega}{\sqrt{2}\omega_i}a_i, \\
P_{2i}
&=&  \frac{1}{2} 
\left(\frac{g_i}{\omega_i}- 1\right) S_{2i}+\frac{1}{2} 
\left(\frac{g_i}{\omega_i} +1 \right) S_{3i}-\frac{\Omega}{\sqrt{2}\omega_i}a_i,
\end{eqnarray*}
where $a_i$ is the cavity photon annihilation operator for the $i$th cavity, and 
\begin{eqnarray*} 
S_{2i}&=&\frac{1}{\sqrt{N_{ai}}}\sum_{j=1}^{N_a}\vert 1\rangle_{jj}\langle 2\vert, \\
S_{3i}&=&\frac{1}{\sqrt{N_{ai}}}\sum_{j=1}^{N_a}\vert 1\rangle_{jj}\langle 3\vert,
\end{eqnarray*}
are the collective atomic operator, with $N_{ai}$ being the total atom number, in the $i$th cavity. For simplicity, we have taken $\Omega_{24}=\Omega_{34}=\sqrt{2} \Omega$, $g_i = \sqrt{N_{ai}}\,g_{14}$, and $\omega_i = \sqrt{g_i^2+\Omega^2}$.
The total Hamiltonian of the system reads:
\begin{equation}
H=\sum_{i}H_i+\sum_{\langle i,j\rangle}H_{\langle i,j\rangle}^t,\\
\end{equation}
where
\begin{equation}
\begin{split}
&H_i=\frac{V_1}{2}P^{\dagger2}_{1i} P^2_{1i} +\frac{V_2}{2}P^{\dagger2}_{2i} P^2_{2i} +UP^{\dagger}_{1i} P_{1i} P^{\dagger}_{2i} P_{2i}\\
&+T^+[P^{\dagger}_{1i}(P^{\dagger}_{1i} P_{1i} +P^{\dagger}_{2i} P_{2i})P_{2i} +P^{\dagger}_{2i}(P^{\dagger}_{1i} P_{1i} +P^{\dagger}_{2i} P_{2i})P_{1i}]\\
&+T^-[P^{\dagger}_{1i} (P^{\dagger}_{1i} P_{1i} -P^{\dagger}_{2i} P_{2i} )P_{2i} +P^{\dagger}_{2i} (P^{\dagger}_{1i} P_{1i} -P^{\dagger}_{2i} P_{2i} )P_{1i}],
\end{split}
\label{eq:ne1}
\end{equation}
represents the Hamiltonian of a two-component polariton in the $i$th cavity, the derivation of which can be found in our previous work \cite{etcbhm1}, and \begin{equation}
H_{\langle i,j\rangle}^t=-t\{P^{\dagger}_{1i}P_{1j} +P^{\dagger}_{2i}P_{2j}\}\,,\label{tunnel}
\end{equation}
describes the tunneling of polaritons between adjcent cavities~\cite{paction2}.
Here, $P^\dagger_{1(2)i}$ and $P_{1(2)i}$ obey bosonic commutation relation: $[P_{\alpha i},\ P_{\beta j}]=0,\ [P^\dagger_{\alpha i},\ P^\dagger_{\beta j}]=0$, and $[P_{\alpha i},\ P^\dagger_{\beta j}]=\delta_{\alpha,\beta}\delta_{i,j}$. As shown in Ref.~\cite{etcbhm1}, the key parameters in Eq.~(\ref{eq:ne1}), $V_1$, $V_2$, $U$ and $T^\pm$, can be tuned over a large extent by appropriately controlling the laser intensities and frequencies. Here to avoid instability, we only consider repulsive interactions such that $V_{1,2}$ and $U$ are all positive. In addition, if we simultaneously change the signs of $T^\pm$, the physics remains unchanged as that sign change can be absorbed by a redefinition of the polariton modes. Hence we will only consider the case with $T^+>0,\ T^-<0$ for simplicity.

In comparison to the single-component BHM, for which there are plenty of theoretical and experimental investigations, our model contains three new key parameters: the inter-species on-site interaction characterized by the interaction strength $U$, the two nonlinear coupling terms characterized by the coupling strength $T^\pm$. Our work will focus on elucidating the effects of these terms.

%
%
%

\section{Mean field phase diagram}

Based on the mean field decoupling theory~\cite{misf1}, we introduce the superfluid order parameters: $\phi_{1(2)}=\langle P_{1(2),i}\rangle$. By the approximation \[ P^\dagger_{1(2),i}P_{1(2),j}=\phi_{1(2)}P_{1(2),j} +\phi^\ast_{1(2)}P^\dagger_{1(2),i}-\phi^\ast_{1(2)}\phi_{1(2)} \,,\] the total Hamiltonian can be decoupled into the following form:
\begin{equation}
\begin{split}
\mathcal{H}_0=&-\mu(P^{\dagger}_{1} P_{1} +P^{\dagger}_{2} P_{2})+\frac{V_1}{2}P^{\dagger2}_{1} P^2_{1} +\frac{V_2}{2}P^{\dagger2}_{2} P^2_{2}\\
&+UP^{\dagger}_{1} P_{1} P^{\dagger}_{2} P_{2}\\
&+T^+[P^{\dagger}_{1}(P^{\dagger}_{1} P_{1} +P^{\dagger}_{2} P_{2})P_{2} +P^{\dagger}_{2}(P^{\dagger}_{1} P_{1} +P^{\dagger}_{2} P_{2})P_{1}]\\
&+T^-[P^{\dagger}_{1} (P^{\dagger}_{1} P_{1} -P^{\dagger}_{2} P_{2} )P_{2} +P^{\dagger}_{2} (P^{\dagger}_{1} P_{1} -P^{\dagger}_{2} P_{2} )P_{1}]\\
&+zt(\phi^\ast_1 \phi_1 +\phi^\ast_2 \phi_2) \,, \\
\mathcal{H}_t=&-zt(\phi^\ast_1 P_1 +\phi^\ast_2 P_2 +h.c.)\,,
\end{split}
\label{eq:ne3}
\end{equation}
where $\mu$ is the chemical potential and $z=2$ is the number of nearest neighbors. For simplicity, here, the site index is neglected. Since Hamiltonian $\mathcal{H}_0$ keeps local particle number conserved, in principle, we can numerically diagonalize $\mathcal{H}_0$ in a subspace with fixed total number of particles. 

\begin{figure}[h]
\centering
    \includegraphics[width=0.95\columnwidth]{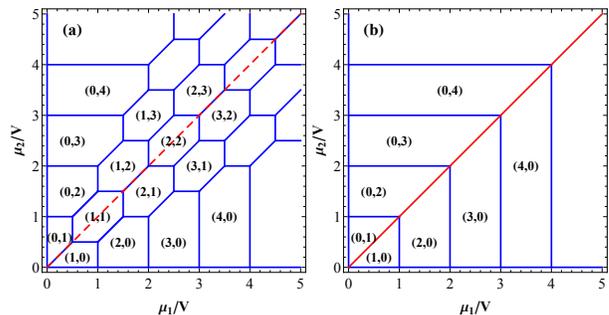}
\caption{(Color online) The ground state of ${\cal H}_0$ with $T^\pm=0$ and $t=0$. Here we have to modify the first term in ${\cal H}_0$ to be $-\mu_1 P^{\dagger}_{1} P_{1} -\mu_2P^{\dagger}_{2} P_{2}$. The values in the bracket indicate those of $n_1$ and $n_2$, respectively. We have used $V_1=V_2=V$, and the value of $U$ is taken as $U=0.5$ in (a), and $U=1.5$ in (b). In all our figures, the quantities $U$, $\mu$, and $T^\pm$ are expressed in units of $V$.}
\label{figphasenew}
\end{figure}

As a reference, let us first set the nonlinear coupling strength $T^\pm=0$, and the tunneling rate $t=0$. Under this situation ${\cal H}_0$ describes a two-component Bose gas with density-density interactions, and it preserves the number of polaritons in each component. The eigenstates therefore correspond to Fock states $|n_1,n_2 \rangle$ with definite integer values of $n_1$ and $n_2$, and with corresponding ground state energy $E(n_1,n_2)$, where $n_\alpha$ ($\alpha=1$, 2) represents the number of polaritons in component $\alpha$ in each cavity. Note that due to the conservation of polariton numbers in individual component, we need to introduce two chemical potentials $\mu_1$ and $\mu_2$ for the two polariton modes. The ground state is determined by the relative size of $V_1$, $V_2$ and $U$. In Fig.~\ref{figphasenew} the ground state phase diagram is shown for the case with equal intra-species interaction strength, i.e., $V_1=V_2=V$. If we focus on the situation with equal chemical potential ($\mu_1=\mu_2=\mu$), it is quite straightforward to show that if the inter-species interaction is smaller than the intra-species one (i.e., $U<V$, see Fig.~\ref{figphasenew}(a)), then the we have a single ground state with $n_1=n_2$ for even total particle number, and two degenerate ground states with $n_1-n_2=\pm 1$ for odd total particle number. By contrast, if $U>V$ (see Fig.~\ref{figphasenew}(b)), no matter the particle number is even or odd, we have two degenerate ground states with $n_1=0$ or $n_2=0$.

The Fock state $|n_1,n_2 \rangle$ discussed above corresponds to the Mott regime with $\phi_1=\phi_2=0$. As inter-cavity tunneling is turned on, $\phi_\alpha$ may take finite values and the system enters the superfluid regime. At zero temperature, the transition from the MI to the SF regimes represents a quantum phase transition. Our goal is to map out a phase diagram by calculating the boundary between these two quantum phases. As we want to focus on the effects of the terms characterized by $U$ and $T^\pm$, for most of our calculation we will choose $V_1=V_2=V$ as the units for energy, and consider the situation with $U < V$ and $U > V$ respectively while $|T^\pm| \ll V$.

\begin{figure}[h]
\centering
    \includegraphics[width=0.95\columnwidth]{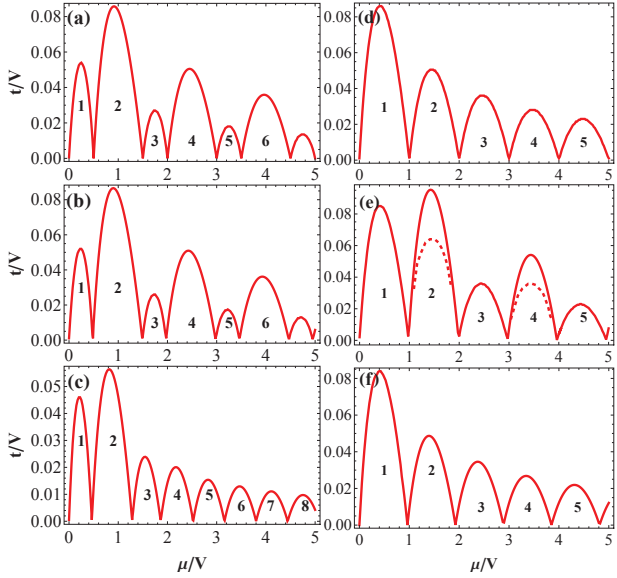}
\caption{(Color online) The mean-field phase diagram where the line represents the boundary between the MI and the SF regime. The region below the line is the MI regime, and the number indicate the total number of polaritons per cavity, $n_{\rm tot}$. The parameter used are (a) $U=0.5$,  $T^+=T^-=0$; (b) $U=0.5$,  $T^+=0$, $T^-=-0.05$; (c) $U=0.5$,  $T^+=0.05$, $T^-=-0.05$; (d) $U=1.5$,  $T^+=T^-=0$; (e) $U=1.5$,  $T^+=0$, $T^-=-0.05$; (f) $U=1.5$,  $T^+=0.05$, $T^-=-0.05$. All the quantities are in units of $V_1=V_2=V$.}
\label{figphase}
\end{figure}

Figure \ref{figphase} shows six examples of the phase diagram obtained from the mean-field decoupling approach with $U<V$ in Fig.\ref{figphase}(a$\sim$c) and $U>V$ in Fig.\ref{figphase}(d$\sim$f). We have also checked that essentially the same phase diagram can be obtained using the Gutzwiller method. The details of the phase diagram will be described in the following.

\subsection{Small Inter-species Interaction ($U<V$)}

In Fig.~\ref{figphase}(a), the nonlinear coupling term is turned off, i.e., $T^\pm=0$. We again focus on the situation with equal chemical potential ($\mu_1=\mu_2=\mu$). Here, the MI regime exhibits an \textquotedblleft even-odd effect\textquotedblright, i.e., the MI region with odd occupation (the number of polaritons per cavity $n_{\rm tot}=2n+1$) is smaller than its nearest neighbors with $n_{\rm tot}=2n$ and $2(n+1)$. A qualitative interpretation of this effect can be provided by studying the excitation gap of a MI \cite{ebhm6}. For even occupation MI with $n_{\rm tot}=2n$, the excitation gap is given by \[ \tilde{\Delta}_{2n}=E(n,n+1)+E(n-1,n)-2E(n,n)=V\,,\] while that for $N_{\rm tot}=2n+1$ is given by \[\tilde{\Delta}_{2n+1}=E(n,n)+E(n+1,n+1)-2E(n,n+1)=U\,.\] Since $V>U$, MI regime with even occupation will be more robust and hence exists in a larger parameter space. The excitation gap will compete with the tunneling energy $-t\sum_{\alpha=1,2}\langle P^\dagger_{\alpha,i}P_{\alpha,i+1}\rangle$ which can be effectively strengthened if there exists degeneracy. As the filling number increases, the tunneling energy grows and the MI regimes with large occupation number become smaller and smaller. However, when the nonlinear coupling terms are present, i.e., $T^\pm \neq 0$, this ``even-odd effect" is weakened and may disappear completely as shown in Fig.~\ref{figphase}(c). An interpretation of $T^\pm$ effect is provided by perturbative analysis in Sec.~IV.  Furthermore, the nonlinear coupling terms could change the degeneracy of the ground state which will be discussed in detail as follows.

\begin{figure}[h]
\centering
    \includegraphics[width=0.95\columnwidth]{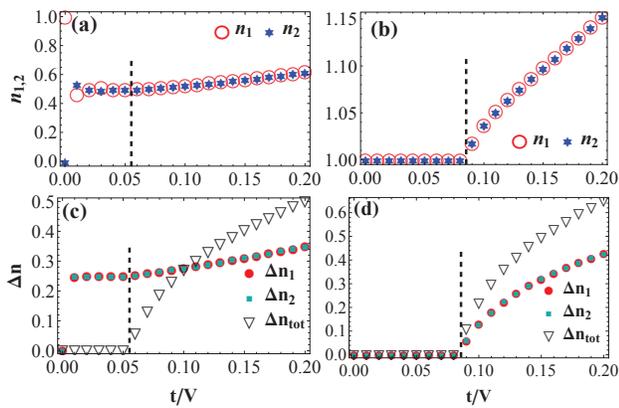}
\caption{(Color online) The average particle numbers and their fluctuations obtained by the Gutzwiller method. For all the subplots, we have $V_1=V_2=V$, $U=0.5$, and $T^\pm=0$. The chemical potential is taken as $\mu=0.25$ for (a) and (c); $\mu=1$ for (b) and (d). The vertical dashed lines indicate the critical tunneling rate $t_c$ beyond which the system changes from MI to SF.}
\label{figGutzwiller1}
\end{figure}

As mentioned previously, if $t$ and $T^\pm$ are all zero, the ground states are doubly degenerate when $n_{\rm tot}=n_1+n_2$ is odd. If $t$ is switched on, due to the super-exchange interaction induced by the inter-cavity tunnelling, the degeneracy would be lifted up and the ground state becomes the superposition state of $n_1-n_2=\pm1$ with equal weight \cite{tcbhm4}. This is indicated in Fig.~\ref{figGutzwiller1}, where we have defined the average particle numbers per cavity $n_\alpha=\langle P^\dagger_\alpha P_\alpha \rangle \ (\alpha=1,2)$ and the fluctuations $\Delta n_\alpha=\langle (P^\dagger_\alpha P_\alpha)^2\rangle -\langle P^\dagger_\alpha P_\alpha\rangle^2$, $\Delta n_{\rm tot}=\langle (P^\dagger_1 P_1+P^\dagger_2 P_2)^2\rangle -\langle P^\dagger_1 P_1+P^\dagger_2 P_2\rangle^2$. For Fig.~\ref{figGutzwiller1}(a) and (c), we take a relatively small chemical potential with $\mu/V=0.25$, for which the average particle numbers are $n_1=n_2=0.5$. In the absence of the tunneling, the ground state is doubly degenerate and are represented by the Fock states $|1,0 \rangle$ and $|0, 1 \rangle$. For $0<t<t_c$, where $t_c$ is the critical tunneling strength beyond which the system changes from MI to SF and is indicated in the figure by the vertical dashed line, the fluctuations of the total number per cavity $\Delta n_{\rm tot}$ vanishes, while $\Delta n_1$ and $\Delta n_2$ are both finite, as shown in Fig.~\ref{figGutzwiller1}(c). Therefore, we can conclude that the ground state is an equal-weight superposition state of $|1,0\rangle$ and $|0,1\rangle$. For the case depicted in Fig.~\ref{figGutzwiller1}(b) and (d), we used a larger chemical potential $\mu/V= 1$ and the average particle number is $n_{\rm tot}=2$. In this case, the MI state is represented by the Fock state $|1,1 \rangle$ with vanishing fluctuations in both $\Delta n_{\rm tot}$ and $\Delta n_{1,2}$. 

\begin{figure}[h]
\centering
    \includegraphics[width=0.95\columnwidth]{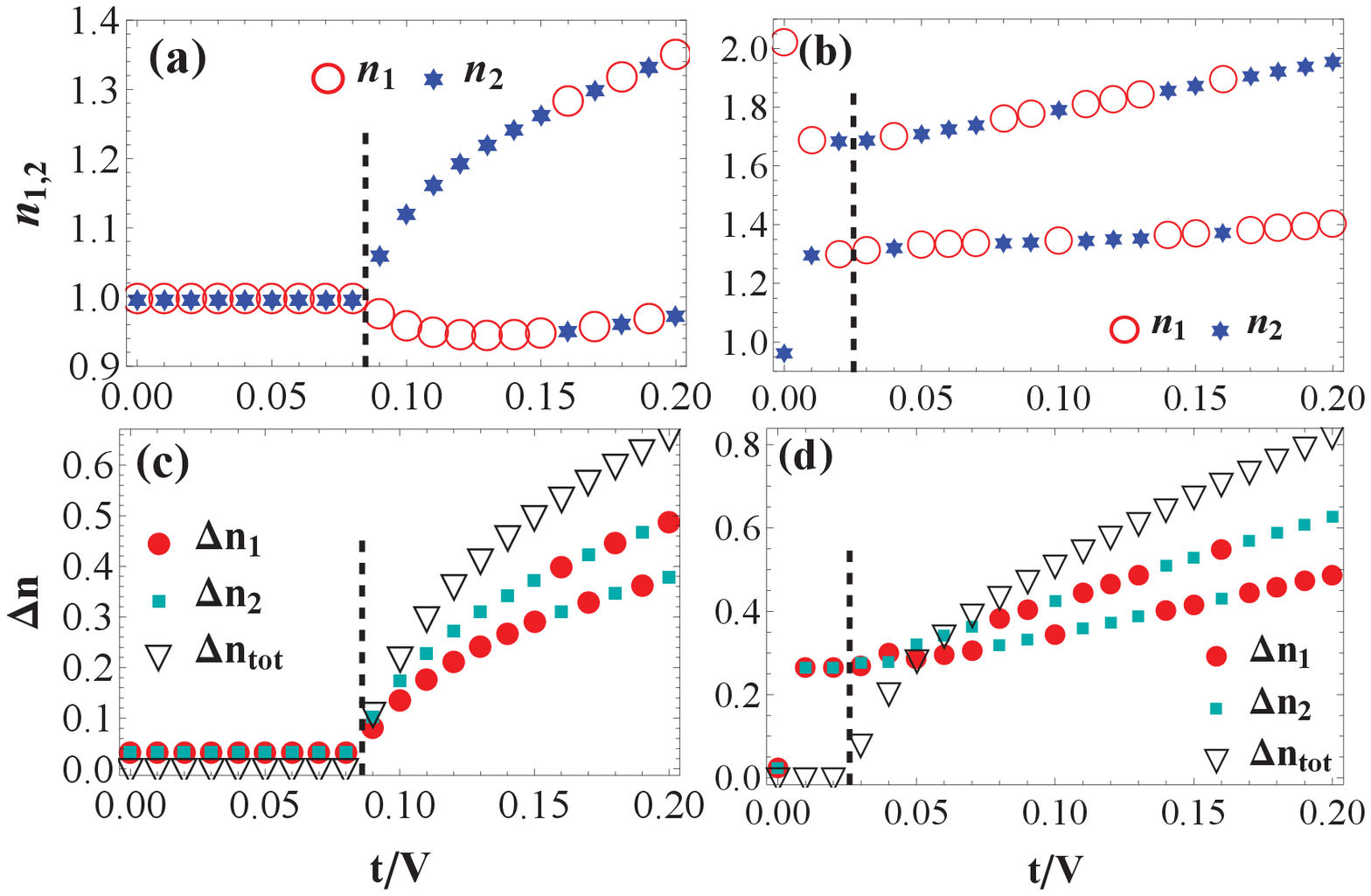}
\caption{(Color online) The average particle numbers and their fluctuations obtained by the Gutzwiller method. For all the subplots, we have $V_1=V_2=V$, $U=0.5$, $T^+=0$, and $T^-=-0.05$. The chemical potential is taken as $\mu=1$ for (a) and (c); $\mu=1.75$ for (b) and (d). The vertical dashed lines indicate the critical tunneling rate $t_c$ beyond which the system changes from MI to SF.}
\label{figGutzwiller2}
\end{figure}

If $T^-$ is turned on, the ground state for $t<t_c$ would be a superposition state Fock states $| n_1, n_2\rangle$ with fixed $n_1+n_2=n_{\rm tot}$. This state is non-degenerate if $n_{\rm tot}$ is even, and can be written as 
\begin{equation} 
|G\rangle=\sum_{n_1,n_2}f_{n_1,n_2}|n_1,n_2\rangle \,, \label{nond}
\end{equation} 
with $f_{n_1,n_2}=(-1)^{\frac{|n_1-n_2|}{2}}f_{n_2,n_1}$. An example is represented in  Fig.~\ref{figGutzwiller2}(a) and (c) with $n_{\rm tot}=2$. For $t<t_c$, we note that the particle number difference remains zero and $n_{1,2}$ have finite fluctuations while $\Delta n_{tot}$ vanishes, so the ground state is a superposition of states $|1,1\rangle$, $|0,2\rangle$ and $|2,0\rangle$. By contrast, when $n_{tot}$ is odd (see Fig.~\ref{figGutzwiller2}(b) and (d)), $T^-$ separates the whole Hilbert space into two degenerate subspaces with $n_1>n_2$ and $n_1<n_2$, respectively, and the corresponding ground states have the form \begin{equation}
|G_+\rangle=\sum_{n_1>n_2}c^+_{n_1,n_2}|n_1,n_2\rangle \,,\label{d1}
\end{equation} 
and 
\begin{equation} |G_-\rangle=\sum_{n_1<n_2}c^-_{n_1,n_2}|n_1,n_2\rangle \,,\label{d2}
\end{equation} 
with $c^-_{n_1,n_2}=(-1)^{\frac{n_2-n_1-1}{2}}c^+_{n_2,n_1}$. One can readily verify that the $T^-$ term has vanishing matrix element between $|G_+ \rangle$ and $|G_- \rangle$. Thus the double degeneracy of the ground state is preserved even if $T^-$ is turned on, and the ``even-odd effect" remains, as shown in Fig.~\ref{figphase}(b). 

However, if we turn on $T^+$, the degeneracy will be lifted, as the $T^+$ term could couple the two degenerate subspaces, i.e., it has non-vanishing matrix elements between $|G_+ \rangle$ and $|G_- \rangle$. This explains the weakening of the ```even-odd effect" as shown in Fig.~\ref{figphase}(c).

\subsection{Large Inter-species Interaction ($U>V$)}

\begin{figure}[h]
\centering
    \includegraphics[width=0.95\columnwidth]{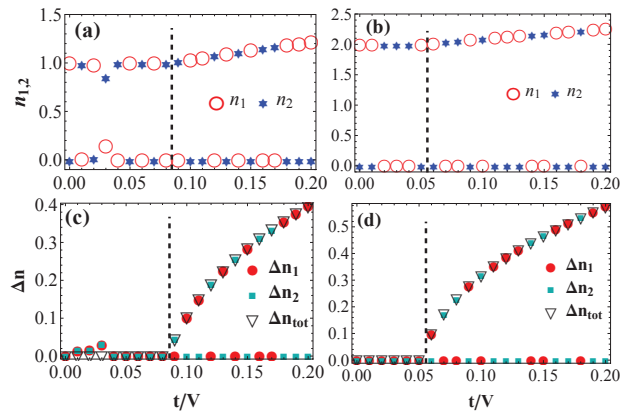}
\caption{(Color online) The average particle numbers and their fluctuations obtained by the Gutzwiller method. For all the subplots, we have $V_1=V_2=V$, $U=1.5$, and $T^\pm =0$. The chemical potential is taken as $\mu=0.5$ for (a) and (c); $\mu=1.5$ for (b) and (d). The vertical dashed lines indicate the critical tunneling rate $t_c$ beyond which the system changes from MI to SF.}
\label{figGutzwiller3}
\end{figure}

In Fig.~\ref{figphasenew}(b), we have known that the ground states, i.e., $|0,n_{\rm tot}\rangle$ and $|n_{\rm tot},0\rangle$, are doubly degenerate for both the even and the odd occupation when $U>V$, $T^\pm=0$ and $\mu_1=\mu_2=\mu$ at $t=0$. In Fig.~\ref{figGutzwiller3}, we plot the occupation numbers and their fluctuations as functions of tunneling rate $t$. we can see that, in the MI region ($0<t<t_c$), the population difference is $n_1-n_2=\pm n_{\rm tot}$ while the number fluctuations $\Delta n_{\rm tot}$ and $\Delta n_{1,2}$ are all zero. It tells us that the ground state is either $|0,n_{\rm tot}\rangle$ or $|n_{\rm tot},0\rangle$, regardless whether $n_{\rm tot}$ is even or odd. In addition, the excitation gap is independent of $n_{\rm tot}$, i.e., $\tilde{\Delta}=V$. As a result, there is no ``even-odd effect" as can be seen from Fig.~\ref{figphase}(d). 

\begin{figure}[h]
\centering
    \includegraphics[width=0.95\columnwidth]{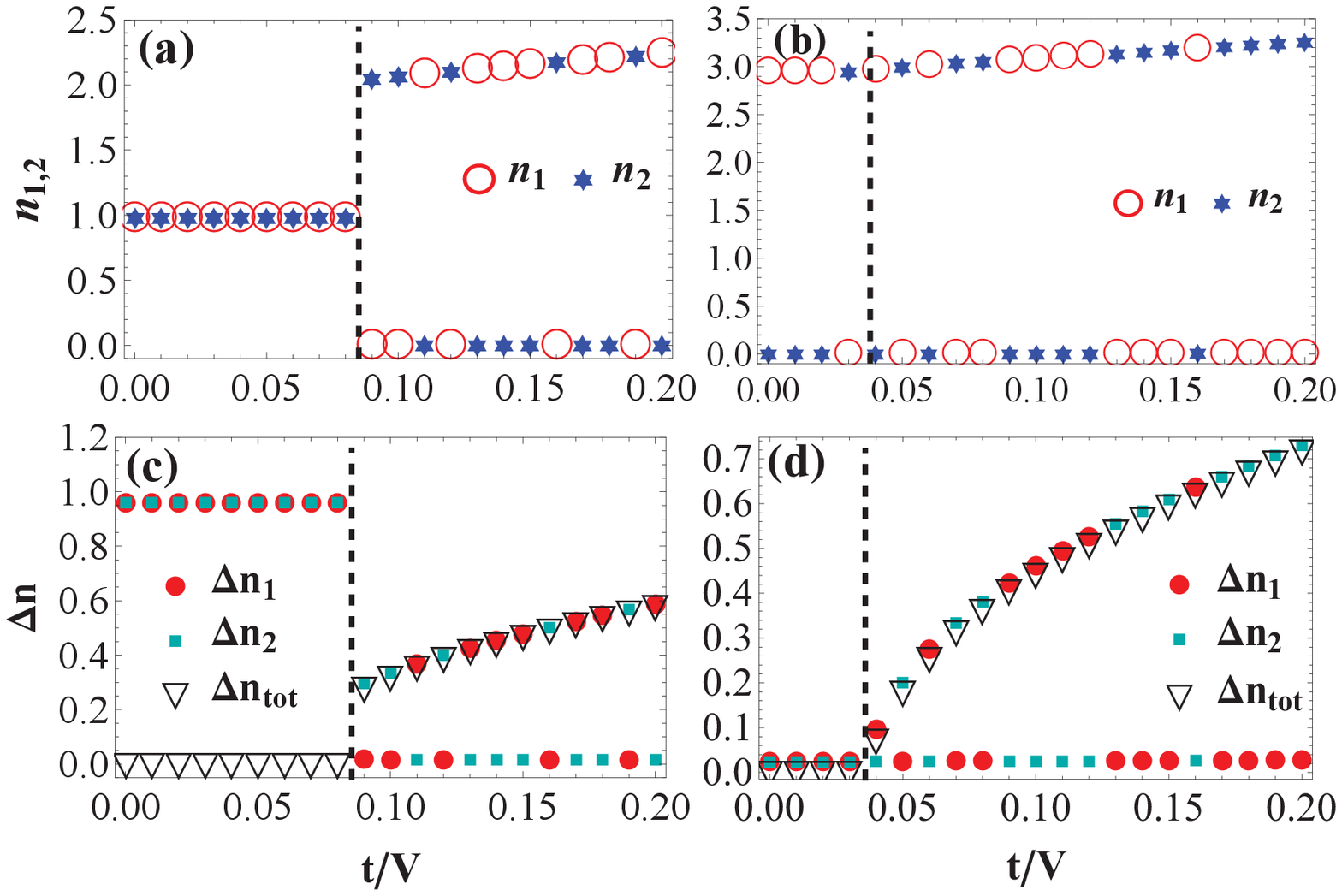}
\caption{(Color online) The average particle numbers and their fluctuations obtained by the Gutzwiller method. For all the subplots, we have $V_1=V_2=V$, $U=1.5$, $T^+=0$, and $T^-=-0.05$. The chemical potential is taken as $\mu=1.5$ for (a) and (c); $\mu=2.5$ for (b) and (d). The vertical dashed lines indicate the critical tunneling rate $t_c$ beyond which the system changes from MI to SF.}
\label{figGutzwiller4}
\end{figure}

However, when $T^-$ is turned on, the ``even-odd effect" emerges as can be seen in Fig.~\ref{figphase}(e). This is because for finite $T^-$, the ground state in the Mott insulator region is non-degenerate when $n_{\rm tot}$ is even, and takes a similar form as in Eq.~(\ref{nond}). An example is shown in Fig.~\ref{figGutzwiller4}(a) and (c). By contrast, for odd $n_{\rm tot}$, $T^-$ can preserve the ground state's double degeneracy and the two degenerate ground states take similar forms as in Eqs.~(\ref{d1}) and (\ref{d2}), but with $c^-_{n_1,n_2}=(-1)^{n_1}c^+_{n_2,n_1}$. It is easy to find that the tunneling energy in this nondegenerate state is much lower than that in $|0,n_{\rm tot}\rangle$ or $|n_{\rm tot},0\rangle$ while the excitation gap decrease is very small. Thus the MI region with even occupation is enlarged by $T^-$ and the ``even-odd effect" reappears. 

Finally, Fig.~\ref{figphase}(f) shows that the ``even-odd effect" vanishes again if $T^+$ is also turned on. The explanation for this is quite similar to the case when $U<V$: the $T^+$ term lifts the degeneracy for odd $n_{\rm tot}$ as its matrix element between $|G_+ \rangle$ and $|G_- \rangle$ is non-zero. 

We comment in passing that, from Fig.~\ref{figphase}(a)$\sim$ (f), we may notice that the MI region with $n_{\rm tot}=1$ remains unchanged for different values of $T^\pm $. This is because the $T^\pm$ terms' matrix elements are all zero in this space.

\begin{figure}[h]
\centering
    \includegraphics[width=0.95\columnwidth]{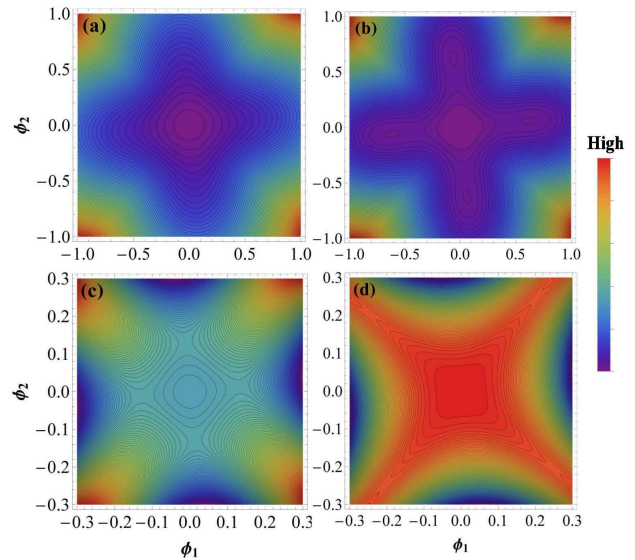}
\caption{(Color online) The energy contours obtained by diagonalizing Eq.~(\ref{eq:ne3}) for $U=1.5$, $\mu=1.5$, $T^+=0$, and $T^-=-0.05$ (all in units of $V_1=V_2=V$). The tunneling rate is given by (a) $t/V=0.05$; (b) $t/V=0.6$; (c) $t/V=0.08$; (d) $t/V=0.1$. Though the order parameters, $\phi_{1,2}$, could also be complex numbers, their relative phase must be $0$ or $\pi$ in the ground state of Eq.~(\ref{eq:ne3}). Hence we can perform the calculations assuming $\phi_{1,2}$ are real, without loss of generality.}
\label{figfirstorder}
\end{figure}

It is not difficult to notice from Fig.\ref{figGutzwiller4}(a) and (c) that there is a discontinuous jump in both the particle numbers and their fluctuations at the critical tunneling rate $t_c$. The jump in $\Delta n_{\rm tot}$, in particular, indicates that the phase transition from MI to SF may be of first order for even $n_{\rm tot}$ when $U>V$ and $T^-\neq 0$. To verify this, we fix $U/V=1.5$, $T^+=0$, $T^-/V=-0.05$, $\mu/V=1.5$, and plot the ground state energy $E(\phi_1,\phi_2)$ with different $t$ in Fig.~\ref{figfirstorder}. When $t$ is very small, $\phi_1=\phi_2=0$ (MI) is the global minimum point (see Fig.~\ref{figfirstorder}(a)). As $t$ increases, additional local minimum points with finite $\phi_{1,2}$ (metastable SF) arise, as shown in Fig.~\ref{figfirstorder}(b). If we continue to increase $t$, Fig.~\ref{figfirstorder}(c) displays that the MI state with $\phi_{1,2}=0$ becomes a local minimum point (i.e., a metastable MI). Finally, the local minimum point becomes a local maximum point if $t$ is sufficiently large  as shown in Fig.~\ref{figfirstorder}(d). In conclusion, metastable SF and metastable MI states  exist near the boundary of the MI lobes in Fig.~\ref{figphase}(e) (between the dashed and solid line) while $n_{\rm tot}$ is even. This existence of such metastable states is a tell-tale signature that the MI-SF phase transition in this case is of first-order. A similar situation can be found in spin-1 bosons which can also host the first-order phase transition \cite{ebhm4,ebhm5}, but they are induced by different effects. In Fig.~\ref{figfirstorder}, another feature is also worth our attention. Only one of the two species is dominant in the SF state, i.e. $|\phi_1|\gg |\phi_2|$ or $|\phi_2|\gg |\phi_1|$ and they are degenerate. This is in consistent with Fig.~\ref{figGutzwiller4}(c) and (d) which indicate that only one component's fluctuation is prominent while the other one's is close to zero. In comparison, both components have significant occupation fluctuations in SF state when $U<V$, as shown in Fig.~\ref{figGutzwiller1}(c) and (d), and Fig.~\ref{figGutzwiller2}(c) and (d).

\begin{figure}[h]
\centering
    \includegraphics[width=0.95\columnwidth]{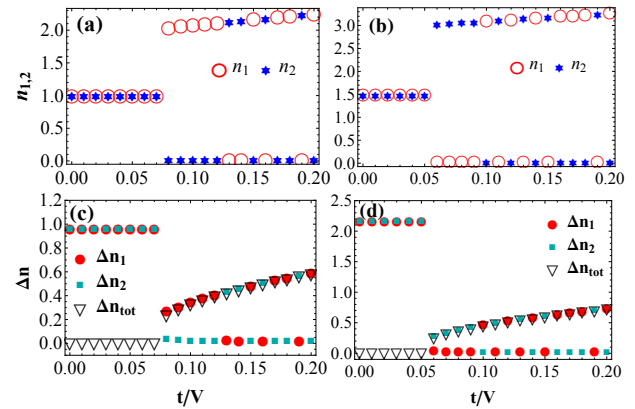}
\caption{(Color online) 
The average particle numbers and their fluctuations obtained by the Gutzwiller method. For all the subplots, we have $V_1=V_2=V$, $U=1.5$, $T^+=0.05$, and $T^-=0$. The chemical potential is taken as $\mu=1.5$ for (a) and (c); $\mu=2.5$ for (b) and (d). }
\label{figfirstorder1}
\end{figure}

Furthermore, if $T^-=0$ but $T^+\neq 0$, not only the even occupation's but also the odd occupation's MI-SF phase transition will be of first-order, as shown in Fig.~\ref{figfirstorder1}. Because all the MI ground states' degeneracy is lifted in the presence of $T^+$ regardless whether $n_{\rm tot}$ is even or odd.

\section{Perturbative Analysis}

\begin{figure}[h]
\centering
    \includegraphics[width=0.95\columnwidth]{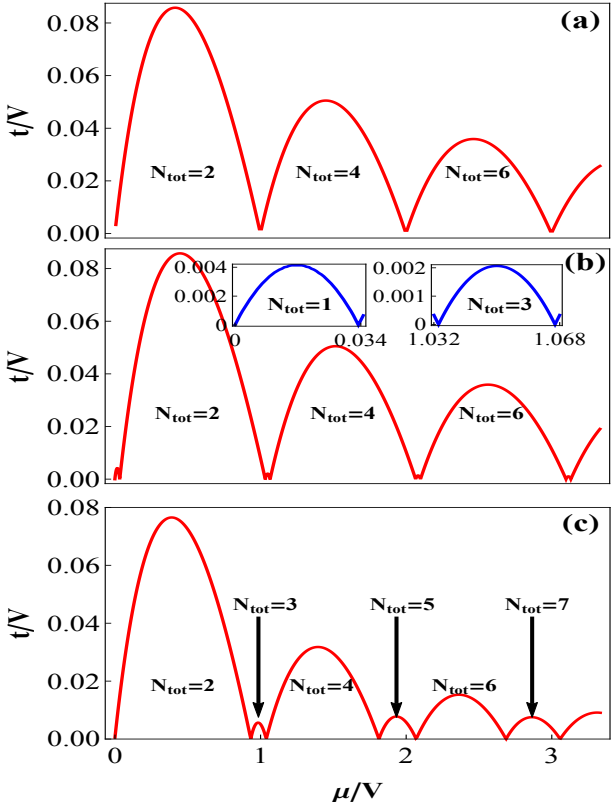}
\caption{(Color online) The mean field phase diagrams for different nonlinear pertuibative terms. Here, $V_1=V_2=V$. (a) phase diagram with $U=0$ and $T^\pm=0$; (b) phase diagram with $U=V/30$ and $T^\pm=0$; (c) phase diagram with $T^+=V/60$ and $U=T^-=0$. }
\label{figMISF}
\end{figure}

To gain more physical insights into the phase diagram, we consider the inter-component interaction (the $U$-term) and the nonlinear coupling (the $T^\pm$-terms) as perturbations and examine how they affect the phase diagram. Furthermore, in order to distinguish the contribution from each of them, we have shown how each of these nonlinear perturbations affect the mean-field phase diagrams in Fig.~\ref{figMISF}.

In Fig.~\ref{figMISF}(a), we show the ``unperturbed" phase diagram by taking $U=0$ and $T^\pm=0$. Here we simply have two uncoupled single-component Bose-Hubbard model. Since we have chosen the intra-component interaction strength $V_1=V_2=V$, the two individual phase diagrams completely overlap with each other. In particular, each MI region is characterized by equal number of the two polariton modes, and hence only even $N_{\rm tot}$ MI regions are present. Along the horizontal axis ($t=0$), the MI region with $N_{\rm tot}=2n$ occupies a region between $(n-1)V<\mu<nV$. At the boundary $\mu=nV$, the four Mott states $|n,n \rangle$, $|n,n+1 \rangle$, $|n+1,n \rangle$ and $|n+1,n+1 \rangle$ are energetically degenerate with energy $-n(n+1)V$.

In Fig.~\ref{figMISF}(b), we show the phase diagram in the presence of a small inter-component intereaction $U \ll V$. As one can see, the $U$-term induces MI regions with odd $n_{\rm tot}=2n+1$, which for small $U$, occurs near the boundaries between two even MI regions at $\mu=nV$. At $t=0$, these odd MI regions are represented by two degenerate Fock states $|n,n+1 \rangle$ and $|n+1,n \rangle$, and occupy a region with $\mu \in (\mu_<, \mu_>)$. The values of $\mu_<$ and $\mu_>$ can be readily obtained from $E(n,n,\mu_<)=E(n,n+1,\mu_<)$ and $E(n,n+1,\mu_>)=E(n+1,n+1,\mu_>)$, from which we obtain the width of the odd MI region in the $\mu$-axis as $\mu_>-\mu_<=U$, which is exactly the excitation gap $\tilde{\Delta}_{2n+1}$ of the MI state. The above analysis provides a more quantitative argument for the even-odd effect mentioned earlier. Furthermore, since the tunneling energy is roughly proportional to $\sqrt{n(n+1)}$, the MI regimes with occupation number $2n+1$ become smaller as $n$ grows. Our mean-field numerical results, as shown in the inset of Fig.~\ref{figMISF}(b), are in complete agreement with this analysis.

In Fig.~\ref{figMISF}(c), we show the phase diagram in the presence of the nonlinear coupling $T^+$-term, while keeping $U=0$ and $T^-=0$. In this case, in the absence of tunneling, the conservation of the number of polaritons in each component is broken, while the total polariton number $n_{\rm tot}$ for each cavity remains conserved for Hamiltonian $H_0$. We also observe the appearance of odd MI regions except for $n_{\rm tot}=1$. Furthermore, the odd MI regions grow in size as $n_{\rm tot}$ increases, which is in stark contrast with the previous case. These properties can be intuitively understood as follows. Consider an odd MI region with $n_{\rm tot}=2n+1$ at $t=0$, which is characterized by two degenerate Fock states $|n,n+1 \rangle$ and $|n+1,n \rangle$ in the absence of $T^+$. When a small $T^+$ is turned on, a direct coupling between these two states is induced with the corresponding matrix element given by $2n(n+1)T^+$. To first order in $T^+$, the ground state energy is shifted down by this amount. The presence of $T^+$-term will also lower the energies of even MI states, but the energy shift is quadractic in $T^+$ as the unperturbed even MI states are nondegenerate. This explains the appearance of odd MI regions with $n_{\rm tot}=2n+1$ for $n>0$. There is no MI region for $N_{\rm tot}=1$, because transition matrix element between Fock states $|0,1 \rangle$ and $|1,0 \rangle$ vanishes. In addition, it is not difficult to show that the width of the odd MI region on the $\mu$-axis is \textbf{$O(4n(n+1)T^+)$}, which grows when $n$ increases, in good agreement with the numerical results.

Actually, no matter $U$ is zero or not, we can always use the above picture, i.e., first order energy shift to the state with degeneracy and second order energy shift to the state without degeneracy, to understand the effect of $T^+$. And it also applies to the situation even $U>V$. As mentioned previously, the ground states with odd occupations are doubly degenerate when $U>V$, $T^-\neq 0$ at $t=0$, and $T^+$ has nonvanishing matrix elements between them which can induce a first order energy shift. At the same time, the energy shift is second order in the even occupation case. On the other hand, the tunneling energy grows as $T^+$ increases. Therefore, the ``even-odd effect" disappears in Fig.~\ref{figphase}(f).

\begin{figure}[h]
    \includegraphics[width=0.9\columnwidth]{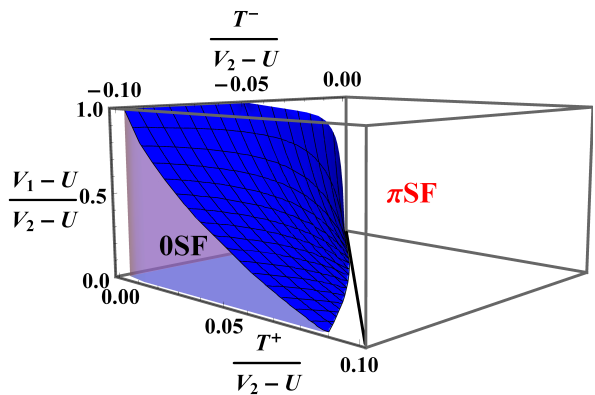}
\caption{(Color online) The phase diagram for the first order phase transition shown in the 3D parameter space ($\frac{T^+}{V_2-U}$, $\frac{T^-}{V_2-U}$, $\frac{V_1-U}{V_2-U}$). The blue surface is the phase boundary, while the 0SF and $\pi$SF is below and above it respectively.}
\label{figPD}
\end{figure}

Finally, let us consider the effect of the $T^-$-term. In the case $U=0$ and $T^+=0$, the $T^-$-term alone would not induce odd MI regions. This is due to the fact that, unlike the $T^+$-term, this term does not induce a direct coupling between $|n,n+1 \rangle$ and $|n+1,n \rangle$ states, i.e., the corresponding matrix element vanishes. Hence a small $T^-$ does not have any noticeable effects on the phase diagram. However, the $T^-$-term can be regarded as a nonlinear coupling between the two components of the polariton, whose effective sign depends on the population difference. As a result, in the superfluid region, when the relative population $n_1-n_2$ changes sign, the relative phase between the two order parameters $\phi_1$ and $\phi_2$ will change from 0 to $\pi$. As this effect is already present in the single-cavity system which we investigated in detail in Ref.~\cite{etcbhm1}, we do not provide a detailed discussion here. Instead, we just present the phase diagram for the superfluid region in Fig.~\ref{figPD}, in which 0SF and $\pi$SF are superfluid phases with relative phase between $\phi_1$ and $\phi_2$ being 0 and $\pi$, respectively. The transition between 0SF and $\pi$SF is of first order.

The perturbation calculation presented above allows us to better understand the phase diagram shown in Fig.~\ref{figphase}. In particular, we can now explain why the inter-species interaction give rise to the even-odd effect, and how the nonlinear coupling term $T^+$ weakens the even-odd effect.

\section{Exact diagonalization}
\begin{figure}[h]
    \includegraphics[width=\columnwidth]{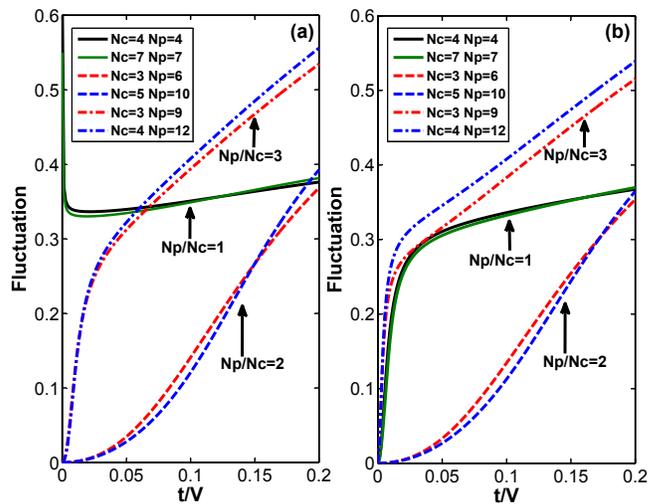}
\caption{(Color online) The fluctuation of the total occupation per cavity $\Delta n$, obtained from the exact diagonalization method. Here $V_1=V_2=V$ and in (a) $U=T^-=0$, $T^+=V/60$; in (b) $U=V/30$, $T^+=T^-=0$. The parameters used in (a) and (b) are the same as those used in Fig.~\ref{figMISF}(c) and (b), respectively.}
\label{figED}
\end{figure}

So far we have investigated a homogeneous system of inter-connected cavity in 1D using a mean-field approach.
In this part, the exact diagonalization method is used to study this model. To make the calculation manageable, we consider a finite number, $N_c$, cavities with $N_p$ total particles. The whole Hilbert space are spanned by the Fock state basis $\vert \psi_i\rangle=\vert n^1_1, n^1_2,\cdots ,n^{N_c}_1, n^{N_c}_2\rangle$ where $n_\alpha^k$ denotes the number of polaritons in component-$\alpha$ in $k$th cavity, and they are constrained as $n^k_{\alpha}\geqslant 0$ and $\sum^{N_c}_{k=1}(n^k_1 +n^k_2)=N_p$. The dimension of the Hilbert space is therefore $D_m=C^{N_p}_{N_p+2N_c-1}$. We write the Hamiltonian into a matrix form under this Fock state basis using the periodic boundary condition, and obtain the ground state ($\vert G\rangle$) of this large sparse matrix through exact diagonalization. In Fig.~\ref{figED}, we plot the total number fluctuation per cavity which is defined as \[\Delta n= \langle G|(n^k)^2 |G \rangle -\langle G|n^k |G \rangle^2 \,\] where $n^k =n_1^k+n_2^k$ is the total number operator for $k$th cavity. Under the periodic boundary condition, this quantity is indepedent of the cavity index $k$. We vary $N_p$ and $N_c$ to some extent while restricting their ratio $N_p/N_c$ (i.e., the number of polaritons per cavity) to be 1, 2, or 3. As can be seen, the behavior of the number fluctuation is sensitive to $N_p/N_c$. The parameters of Fig.~\ref{figED}(a) are the same as those used in Fig.~\ref{figMISF}(c). Here one can see that in the limit $t \rightarrow 0$, the number fluctuation for systems with $N_p/N_c=1$ remains finite, which indicates the lack of MI region for $N_{\rm tot}=1$. By contrast, in the same limit, the number fluctuations for $N_p/N_c=2$ and 3 vanish and more specifically $\Delta n$ for $N_p/N_c=3$ tends to zero with a much steeper slope. We thus expect to see a large MI region for $N_{\rm tot}=2$ and a small MI region for $N_{\rm tot}=3$. All these are fully consistent with the mean-field phase diagram shown in Fig.~\ref{figMISF}(c). The parameters used for Fig.~\ref{figED}(b) are the same as those used for Fig.~\ref{figMISF}(b). Here $\Delta n$ for all three ratios or $N_p/N_c$ vanish in the limit $t \rightarrow 0$. The slopes of $\Delta n$ near this limit indicate that two small MI regions for $N_{\rm tot}=1$ and 3, and a large MI region for $N_{\rm tot}=2$ in the mean-field limit is expected, which is again consistent with the results obtained earlier.

\section{Summary}

In summary, we have presented a scheme to realize a two-component BHM with nonlinear inter-component coupling in a system of cavity polaritons. We mapped out the phase diagram showing the boundaries between the MI phase and the SF phase. Using several different approaches --- the mean-field decoupling method, the Gutzwiller method, the perturbation calculation and the exact diagonalization, we show how the inter-species interaction and the nonlinear coupling terms affect the phase diagram, and particularly how they induce the first-order MI-SF phase transition and give rise to or weaken the \textquotedblleft even-odd effect\textquotedblright. Additionally, the competition between the nonlinear coupling strengths $T^+$ and $T^-$ can drive a first order quantum phase transition within the SF regime that changes the relative condensate phase of the two polariton components. Through our study, we have obtained a clear understanding about this two-component BHM. In the future, we could also realize an two-component BHM that breaks the time reversal symmetry \cite{TRS,ebhm8} by manipulating the external fields' relative phase. This model may host more exotic $p$-wave superfluid phases \cite{ebhm7}.

\section{Acknowledgement}

This work was funded by National Basic Research Program of China 2011CB921204,
2011CBA00200, and the "Strategic Priority Research Program(B)" of the Chinese
Academy of Sciences, Grant No. XDB01030200, National Natural Science Foundation
of China (Grant Nos. 11174270, 11474266). Z. -W. Zhou gratefully acknowledges
the support of the K. C. Wong Education Foundation, Hong Kong. HP acknoledges
support from US NSF.

\end{document}